\documentclass[twocolumn,showpacs,prb,amsmath,amssymb,aps]{revtex4}% Physical Review B

\usepackage{graphicx}% Include figure files
\usepackage{dcolumn}% Align table columns on decimal point
\usepackage{bm}% bold math

\begin{document}

%\preprint{APS/123-QED}

\title{Effects of broken time-reversal symmetry on transmission zeros
in the Aharonov-Bohm interferometer}

\author{Tae-Suk Kim}
\affiliation{Institute of Physics and Applied Physics, Yonsei University, 
  Seoul 120-749, Korea}
\author{Sam Young Cho and Chul Koo Kim}
\affiliation{Institute of Physics and Applied Physics, Yonsei University, 
  Seoul 120-749, Korea \\
 Center for Strongly Correlated Materials Research,                           
          Seoul National University, Seoul 151-742, Korea}
\author{Chang-Mo Ryu}
\affiliation{Department of Physics,                                                       
          Pohang University of Science and Technology,          
          Pohang 790-784, Korea}

\date{\today}% It is always \today, today,

\begin{abstract}
 In this paper, we study the behavior of the transmission zeros 
in the closed Aharonov-Bohm(AB) interferometer with an embedded 
scattering center in one arm
and the corresponding change in the transmission phase when 
the time-reversal symmetry is broken by magnetic fields. 
Specifically, we consider three embedded scattering centers:
one discrete energy level, a double-barrier well, and a $t$-stub.
We find the followings from our model study: 
(i) The transmission zeros are real 
when the AB flux is an integer or a half-integer multiple of the flux quantum, 
and the transmission phase jumps by $\pi$ at the zeros. 
(ii) The transmission zeros become complex or 
are shifted off the real-energy axis 
when the magnetic AB flux is not an integer or a half-integer
multiple of the flux quantum, 
and the transmission phase evolves continuously.
(iii) The distance of the zeros from the real-energy axis or the imaginary 
part of the transmission zeros is sinusoidal as a function 
of the magnetic AB phase.
We suggest the experimental setup which can test our results. 
\end{abstract}

\maketitle

\section{Introduction}
 Recent advances in nanotechnology have made it possible 
to measure the phase of electron wave function
\cite{Yacoby,schuster,reflectp,kondophase1,kondophase2}.
In bulk systems, phase coherence of electron wave functions can be washed out by  
inelastic scattering processes. On the other hand, the phase of electron 
wave function can be preserved in nanoscopic systems. Typical experimental 
tools, which can measure the electron's phase, are the Aharonov-Bohm (AB) 
interferometers. To study the phase evolution due to the target system, 
the system is inserted in one of two arms of the AB interferometer. 
The I-V curves are measured between the external electrodes
connected to the AB ring as function of the AB magnetic flux 
while other control parameters, e.g., the Fermi energy level, 
are varied. The phase evolution of the electron wave functions
in the target system is extracted from the measured I-V curves.

 The closed AB interferometer in a two-terminal 
configuration does not yield much information about the phase shift
in the target system due to the phase locking effect\cite{Buttiker1,Buttiker2}. 
Multiple windings of the electron motion 
along the AB ring result in the conductance which is even\cite{Buttiker1,Buttiker2}
 in the AB flux $\Phi$ or $G(-\Phi) = G(\Phi)$.  
This Onsager relation\cite{onsager} constrains 
the measured phase to be either $0$ or $\pi$.
This phase-locking effect can explain the observed phase jump by $\pi$ 
at the conductance peaks\cite{Yacoby}
of the closed AB interferometer with an embedded quantum dot.
The same phase of the transmission amplitude was observed\cite{Yacoby}
at successive Coulomb peaks. This means that an additional phase shift 
by $\pi$ should occur in between two successive Coulomb peaks. The existence of 
the transmission zeros can also explain this feature\cite{hwlee,sycho}.

 In order to observe the phase evolution in the target system, 
an open AB interferometer was devised by Schuster, et. al.\cite{schuster}, 
which is similar to the double-slit experiments. 
Suppression of the back-scattered electrons prevents the multiple 
windings along the AB ring so that the total transmission amplitude 
becomes sum of two transmission amplitudes through the upper and lower arms,  
\begin{eqnarray}
t &=& t_l + t_u e^{i\phi}. 
\end{eqnarray}
Here $\phi$ is the AB phase due to the magnetic flux. 
Measuring the conductance which is proportional to $|t|^2$ as function 
of the AB phase, the transmission phase of the target system 
can be directly obtained.

 In a phase-coherent system, the two different phases can be defined.
The Friedel phase $\theta_f$ is defined as the argument of 
the determinant of the scattering $S$-matrix, $e^{2i\theta_f} = \mbox{det} S$. 
The change in the Friedel phase is 
related to the density of states via the Friedel sum rule\cite{langreth}, 
$\rho(E) = \pi^{-1} d\theta_f / dE$. 
The measured phase in the open AB interferometers is in fact the transmission
phase, the argument of the transmission amplitudes 
($t = |t| e^{i\theta_t}$). 
Recent works\cite{hwlee,buttiker} made clear distinctions between the two phases. 
In the absence of the transmission zeros, the two phases are 
identical. The transmission phase jumps by $\pi$ at 
the real transmission zeros 
and the two phases differ by this amount\cite{hwlee,sycho,buttiker}. 
In a time-reversal symmetric system, 
all the possible transmission zeros are proven to be {\it real}\cite{hwlee,hwlee2}.  
For more details about the transmission phase in the time-reversal symmetric 
case, see the works of Lee\cite{hwlee} and of Taniguchi and B\"{u}ttiker
\cite{buttiker}.

 In this paper, we address the following question:
What will happen to the transmission zeros and the transmission
phase when the time-reversal symmetry is broken by the external fields, e.g., 
the magnetic fields? 
Specifically we answer this question by studying 
the closed AB ring (see Fig.~\ref{abring}) with an embedded scattering center 
in the presence of the magnetic AB flux.
Since the AB ring provides the transmission zeros due to the destructive
interference between two arms and the time-reversal symmetry can be 
broken by applying the magnetic fields, the AB interferometer 
is an ideal system for our purpose. 
In connection to our work, we note that 
the effects of broken {\it unitarity} in the AB ring
on the phase-locking were investigated by other group\cite{israel}.

 In general, either one transmission pole or one transmission zero 
gives rise to the phase change by $\pi$ as the Fermi energy is scanned
through the real part of pole or zero. 
The poles always lie in the lower half-plane of the 
complex-energy plane due to the causality relation. On the other hand,
the zeros can be anywhere in the complex-energy plane as we will discuss below
when the time-reversal symmetry is broken. 
Poles give the same contribution to the Friedel phase $\theta_f$
 and the transmission phase $\theta_t$.
The energy scale over which the phase evolution 
occurs is determined by the imaginary part of the transmission poles. 
On the other hand, transmission zeros give an additional contribution $\theta_z$
only to the transmission phase. 
The transmission phase can be written as the sum of two: 
$\theta_t = \theta_f + \theta_z$. 
Depending on the position of zeros in the complex-energy plane, 
the behavior of the phase evolution becomes quite different. 
When the transmission zeros
lie on the real-energy axis, abrupt phase jump by $\pi$ is observed. 
Varying the AB magnetic flux, the transmission zeros can be shifted 
off the real-energy axis. In this case, the transmission phase 
evolution is continuous and occurs over the energy scale -- 
the imaginary part of transmission zeros -- as the Fermi energy is 
scanned. 
That is, the transmission phase
becomes a discontinuous function of control parameters when 
the transmission zeros hit the real-energy axis.

 To summarize the results of our study, the transmission zeros 
of $t$(the transmission amplitude of the closed AB ring),
based on their position in the complex-energy plane,  
can be grouped into three different classes.
\begin{itemize}
\item Class I: Transmission zeros lie on the real-energy axis. 
 The trajectory of the transmission amplitude $t$ passes through 
 the origin and the transmission phase $\theta_t$ jumps by $\pi$ 
 at the transmission zero. 

\item Class II: Poles and zeros lie in the same lower half-plane of the 2d 
 complex-energy plane. The trajectory does not encircle the origin,
 and the evolution of $\theta_t$ is continuous and its range is 
 confined by two extreme points of 
 the trajectory when viewed from the origin. 
 Each pole and zero give rise to the phase change by $\pi$, but
 the sign is opposite. The combined effect of one pole and one zero
 is a smooth evolution of $\theta_t$ and 
 the difference $\Delta \theta_t$ before and after passing through 
 one zero and one pole approaches $0$ or $\Delta \theta_t = 0$. 

\item Class III: Poles(zeros) lie in the lower(upper) half-plane, 
 respectively.  
 The trajectory encircles the origin. One pole and one zero give 
 the same sign of the phase evolution and $\Delta \theta_t = 2\pi$. 
 
\end{itemize}
Trajectories of the transmission amplitude $t$ and its phase are schematically
shown in Fig.~\ref{3tzero} to help the readers to understand three different
classes of transmission zeros, $Z_z$. 
Depending on the nature of the scattering centers which are inserted
into one arm of the AB interferometer, all three classes or some of them
can be realized by varying magnetic AB flux. 
We find from our study that the transmission zeros are real 
when the AB flux is an integer or a half-integer multiple of the flux quantum, 
and the transmission phase jumps by $\pi$ at the zeros. 
The transmission zeros become complex or 
shifted off the real-energy axis 
when the magnetic AB flux is not an integer or a half-integer
multiple of the flux quantum, 
and the transmission phase evolves continuously.
The distance of the zeros from the real-energy axis or the imaginary 
part of the transmission zeros are sinusoidal as function 
of the magnetic AB phase.

 A general formulation on the $S$-matrix for the AB ring is presented 
in the Appendix~\ref{app_abring} when scattering centers are present 
along the arms of the AB ring. 
These scattering centers and the accrued phase by the 
motion along the AB ring can be parameterized by the matrices
${\cal R}$'s and ${\cal T}$'s as described in the Appendix~\ref{app_abring}. 
In this paper, we are interested in the AB ring (see Fig.~\ref{abring}) 
when the target system is inserted in the lower arm. 
When the target system is described by the scattering matrix 
$S_0 = \left( \begin{smallmatrix} r_0 & t_0' \\ t_0 & r_0' \end{smallmatrix} \right)$,
the matrices ${\cal R}$'s and ${\cal T}$'s are given by the generic forms,
\begin{eqnarray}
{\cal R} &=& z_F \begin{pmatrix} 0 & 0 \cr 0 & r_0 \end{pmatrix}, ~~
 {\cal T} ~=~ z_F \begin{pmatrix} e^{i\phi/2} & 0 
                  \cr 0 & t_0 e^{-i\phi/2} \end{pmatrix}, \nonumber\\
{\cal R}' &=& z_F \begin{pmatrix} 0 & 0 \cr 0 & r_0' \end{pmatrix}, ~~
 {\cal T}' ~=~ z_F \begin{pmatrix} e^{-i\phi/2} & 0 
                   \cr 0 & t_0'e^{i\phi/2} \end{pmatrix}.
\end{eqnarray}
Here $\phi = 2\pi \Phi \cdot e/hc$ is the AB phase due to the 
magnetic flux $\Phi$ passing through the AB ring. 
Half of the AB phase is attached to each of the lower and the upper arms 
of the AB ring. The trajectory of $t$ depends on the chosen gauge or 
the way how the AB phase is inserted into the scattering matrix. 
Of course, the measurable quantities like $|t|^2$ and $\Delta \theta_t$ 
do not depend on the gauge. 
$z_F = e^{ik_FL}$ is the phase
accrued by the motion of electrons along either of two arms of length $L$. 
Since we are interested in a phase-coherent system, 
we restrict our study to $T=0$ K. 
The incident electrons will be confined to the Fermi energy with the wave number
$k_F$ in our study.

 In subsequent sections, we illustrate the effects of broken time-reversal 
symmetry on the transmission zeros in the AB interferometer by studying
three model systems. 
In Sec.~\ref{sectdot}, we consider the 
AB interferometer with one discrete energy level. 
This system is simple enough to 
obtain the $S$-matrix in a closed form and allows one to study analytically
the behavior of the transmission zero under the AB flux. 
In Sec.~\ref{sectdbrt} and \ref{sectstub}, 
we study the AB ring when two different
types of multi resonant level systems are inserted in the lower arm.
In Sec.~\ref{sectdbrt}, we study the transmission properties of the AB ring
with an embedded double-barrier well. The double-barrier well    
provides multi discrete energy levels
through which the resonant tunneling is realized, but 
the transmission probability never becomes zero for this system. 
In Sec.~\ref{sectstub}, a $t$-stub with the double-barrier
is inserted in the AB ring. In contrast to the double-barrier well,  
the $t$-stub accommodates the transmission zeros as well as the multi resonant 
levels. 
Our study is summarized in Sec.~\ref{sectsum}.

\section{\label{sectdot} AB interferometer with an embedded resonant level}
 In this section, we consider the Aharonov-Bohm interferometer shown 
in Fig.~\ref{dotsystem}. 
This model system may be the simplest one which can accommodate 
the zero-pole pair in the transmission amplitude.
This system contains both the direct tunneling between two leads
and the resonant tunneling through one discrete energy level in the dot. 
The study of this simple model system helps us to analyze 
more realistic and complex systems to be discussed in the subsequent sections. 
The transmission amplitude of this system is characterized with one 
pole and one zero. 
The pole is provided by the discrete level in the dot while the zero 
is the result of the destructive interference in the AB ring geometry. 
Three different classes in the trajectories of $t$,
 summarized in the introduction, 
can be all realized with the variation of the AB phase.

 We can derive the scattering matrix of the AB interferometer 
using the $t$-matrix method with 
$S_{ij} = \delta_{ij} - 2\pi \delta(E_i - E_j) T_{ij}$ or 
the Green's function method. 
\begin{eqnarray}
S_{\rm ring} &=& \begin{pmatrix} r_{LL}^{\phantom{*}} & t_{RL}^{\phantom{*}} \cr
               t_{LR}^{\phantom{*}} & r_{RR}^{\phantom{*}} \end{pmatrix}, 
\end{eqnarray}
where the reflection and transmission amplitudes are given by 
the equations,
\begin{subequations}
\begin{eqnarray}
t_{RL}^{\phantom{*}}
 &=& -i \sqrt{T_0} - \overline{\Gamma} G_d^{r} (\epsilon) 
   \left[ \sqrt{T_0} + \sqrt{g} \sin \phi  \right.  \nonumber\\
 &&  \hspace{1.0cm} \left. + i\sqrt{g(1-T_0)} \cos \phi  \right], \\
t_{LR}^{\phantom{*}}
 &=& -i \sqrt{T_0} - \overline{\Gamma} G_d^{r} (\epsilon) 
   \left[ \sqrt{T_0} - \sqrt{g} \sin \phi  \right. \nonumber\\
 &&  \hspace{1.0cm} \left. + i\sqrt{g(1-T_0)} \cos \phi  \right], \\
r_{LL}^{\phantom{*}}
 &=& \sqrt{1-T_0} - \overline{\Gamma} G_d^{r} (\epsilon)  \nonumber\\
 && \times \left[ \frac{2i}{1 + \gamma} - 2i \frac{\Gamma_R}{\Gamma} 
         + \sqrt{gT_0} \cos \phi \right],  \\ 
r_{RR}^{\phantom{*}}
 &=& \sqrt{1-T_0} - \overline{\Gamma} G_d^{r} (\epsilon)  \nonumber\\
 && \times \left[ \frac{2i}{1 + \gamma} - 2i \frac{\Gamma_L}{\Gamma} 
         + \sqrt{gT_0} \cos \phi \right]. 
\end{eqnarray}
\end{subequations}
The angle $\phi$ is the Aharonov-Bohm phase $2\pi \Phi/\Phi_0$, where 
$\Phi$ is the magnetic flux threading through the AB ring and $\Phi_0=hc/e$ 
is the flux quantum. The gauge is chosen such that the AB phase $\phi$ 
is attached to the tunneling matrix $V_{dR}$ as $V_{dR} = |V_{dR}| e^{i\phi}$. 
$T_0 = {4\gamma / (1+\gamma)^2}$ is the direct tunneling probability, 
where $\gamma = \pi^2 N_L N_R |T_{LR}|^2$. $N_L$ and $N_R$ are the density of 
states(DOS) in the left and right leads, respectively. Other parameters 
are defined as $\Gamma_p = \pi N_p |V_{dp}|^2$ ($p=L,R$), 
$\Gamma = \Gamma_L + \Gamma_R$, $\overline{\Gamma} = \Gamma/(1 + \gamma)$, 
and $g = 4\Gamma_L \Gamma_R /\Gamma^2$.

 The $S$-matrix satisfies the Onsager relation, 
$S_{ij}(\phi) = S_{ji}(-\phi)$ under the inversion of the magnetic flux.
If the system is mirror-symmetric under the transformation $L\leftrightarrow R$
or the quantum dot is coupled symmetrically to the left and right leads
($\Gamma_L = \Gamma_R$), we obtain the symmetric relation: 
$r_{LL}^{\phantom{*}} = r_{RR}^{\phantom{*}}$.

 The discrete energy level in a dot is broadened with the linewidth $\Gamma$ 
due to the coupling to the left and right leads.
The direct tunneling between the two leads further renormalizes
the linewidth($\overline{\Gamma}$) and shifts the energy level position. 
The retarded Green's function $G_d^{r}$ of a dot is given by the equation, 
\begin{eqnarray}
G_d^{r} (\epsilon) 
 &=& \frac{1}{\epsilon - \epsilon_d(\phi) + i\overline{\Gamma} }. 
\end{eqnarray}
Here $\epsilon_d(\phi)
 = \epsilon_d - \overline{\Gamma} \sqrt{g\gamma } \cos\phi$
is the renormalized energy level of a dot.
The transmission probability can be readily calculated from 
$T (\epsilon) = |t_{RL}^{\phantom{*}}|^2$, 
\begin{eqnarray}
T (\epsilon) 
 &=& T_0 + 2 \overline{\Gamma} \sqrt{gT_0(1-T_0)} \cos\phi ~ \mbox{Re} G_d^{r} 
    \nonumber\\
 && + \overline{\Gamma} [ T_0 - g(1-T_0\cos^2\phi) ] ~ \mbox{Im} G_d^{r}. 
\end{eqnarray}
The transmission probability $T(\epsilon, \phi)$ is an even function 
of the AB phase $\phi$, satisfying the Onsager relation.

 We may rewrite the elements of the scattering matrix in other forms,
\begin{subequations}
\begin{eqnarray}
t_{LR}^{\phantom{*}}
 &=& -i G_d^{r} (\epsilon) \left[ \sqrt{T_0} (\epsilon - \epsilon_d)       
    + \overline{\Gamma} \sqrt{g} e^{i\phi} \right], \\
\label{trnscofdot}
t_{RL}^{\phantom{*}}
 &=& -i G_d^{r} (\epsilon) \left[ \sqrt{T_0} (\epsilon - \epsilon_d) 
    + \overline{\Gamma} \sqrt{g} e^{-i\phi} \right], \\
r_{LL}^{\phantom{*}}
 &=& G_d^{r} (\epsilon) \left[ \sqrt{R_0} (\epsilon - \epsilon_d)
    - \overline{\Gamma} \sqrt{g\gamma} \cos\phi  \right. \nonumber\\
 && \hspace{1.0cm} \left. 
    + i (\overline{\Gamma}_R - \overline{\Gamma}_L) \right], \\
r_{RR}^{\phantom{*}}
 &=& G_d^{r} (\epsilon) \left[ \sqrt{R_0} (\epsilon - \epsilon_d)
    - \overline{\Gamma} \sqrt{g\gamma} \cos\phi  \right.  \nonumber\\
 && \hspace{1.0cm} \left.  
    - i (\overline{\Gamma}_R - \overline{\Gamma}_L) \right].
\end{eqnarray}
\end{subequations}
Here $R_0 = 1- T_0$ is the reflection probability of the direct tunneling. 
Using the above expressions, we can show 
the unitarity of the $S$-matrix, 
$| t_{LR}^{\phantom{*}} |^2 + | r_{LL}^{\phantom{*}} |^2 = 1$ and 
$r_{LL}^{\phantom{*}} t_{LR}^{*} + t_{RL}^{\phantom{*}} r_{RR}^{*} = 0$.

 The Friedel phase $\theta_f$ can be found from the determinant of $S$
which can be written in terms of the Green's function of a dot, 
\begin{subequations}
\begin{eqnarray}
\mbox{det} S 
 &=& \frac{G_d^{r}} {G_d^{a}} 
 ~=~ \frac{ \epsilon - \epsilon_d(\phi) - i \overline{\Gamma} }
          {\epsilon - \epsilon_d(\phi) + i \overline{\Gamma} }  
 ~=~ e^{2i\theta_f},  \\
\label{Fridelp}
\theta_f 
 &=& \frac{\pi}{2} 
  + \tan^{-1} \frac{\epsilon - \epsilon_d(\phi)}{\overline{\Gamma}}.
\end{eqnarray} 
\end{subequations}
$G_d^{a} = [G_d^{r}]^*$ is the advanced Green's function of a dot. 
The Friedel phase changes smoothly from $\theta_f=0$ to $\theta_f=\pi$ 
as the energy level of a dot is scanned through the Fermi energy. 
%The continuous evolution of $\theta_f$ is not influenced by 
%the transmission zeros. 

 The transmission phase $\theta_t$ is obtained from 
$t = t_{RL}^{\phantom{*}} = |t_{RL}^{\phantom{*}}| e^{i\theta_t}$
and is written as the sum of the Friedel phase 
and the contribution from the zero. 
The Eq.~(\ref{trnscofdot}) can be written as
\begin{eqnarray}
t &=& -i \frac{ (\epsilon - \epsilon_d) \sqrt{T_0}
      + \overline{\Gamma} \sqrt{g} \cos \phi 
      - i \overline{\Gamma} \sqrt{g} \sin \phi  }  
    { \epsilon - \epsilon_d 
      + \overline{\Gamma} \sqrt{g\gamma} \cos \phi + i \overline{\Gamma} }.
\end{eqnarray}
The transmission amplitude at the Fermi energy 
$t(\epsilon=0)$ is plotted in the 2d complex plane 
in Fig.~\ref{ringdot}(a) and \ref{ringdot}(d) 
as an implicit function of $\epsilon_d$
while varying the AB phase $\phi$. 
All the trajectories of $t$ are circles. 
The position of a discrete energy level $\epsilon_d$, 
which can be shifted with the gate voltage capacitatively coupled to the dot,
is varied from the empty state to the filled state. 
That is, the value of $\epsilon_d$ is changed from $\infty$ to $-\infty$.  
Writing $Z = \epsilon - \epsilon_d$, we can rewrite $t=t_{RL}^{\phantom{*}}$ as
\begin{eqnarray}
t &=& -i \sqrt{T_0} \frac{Z - Z_z} {Z - Z_p}, 
\end{eqnarray}
in terms of the pole $Z_p$ and the zero $Z_z$. 
This zero-pole pair is given by the expressions, 
\begin{subequations}
\begin{eqnarray}
Z_p &=& -\overline{\Gamma} \sqrt{g\gamma} \cos \phi - i \overline{\Gamma}, \\
\label{dotzero}
Z_z &=& -\overline{\Gamma} \sqrt{\frac{g}{T_0}} [ \cos\phi - i \sin\phi].
\end{eqnarray}
\end{subequations}
Note that the imaginary part of the transmission zero is proportional to 
$\sin\phi$ and vanishes when $\phi=n\pi$($n$ is an integer). 
%[Elaborate on the structure of the transmission zero]

%Class I 
 When the magnetic AB flux is an integer or a half-integer multiple of
the flux quantum, or when $\phi = n\pi$ ($n$ is an integer), 
the imaginary part of $Z_z$ vanishes and the transmission zero lies 
on the real-energy axis. 
The trajectory of $t$ passes through the origin (the class I). 
The analytic expression of the transmission phase $\theta_t$ when $\phi = n\pi$
is given by the equation,   
\begin{subequations}
\begin{eqnarray}
\theta_t &=& \theta_f + \theta_z, \\
\theta_z &=& \pi 
  - \pi \Theta(\epsilon - \epsilon_d + (-1)^n \overline{\Gamma} \sqrt{g/T_0} ). 
\end{eqnarray}
\end{subequations}
Here $\Theta(x)$ is the step function. 
$\theta_f$ is the Friedel phase given by the Eq.~(\ref{Fridelp})
and $\theta_z$ is the contribution from the transmission zero.  
Since the transmission zero is real, the transmission phase jumps abruptly by $\pi$ 
as shown in Fig.~\ref{ringdot}(b) and (e). 
The zero-pole pair leads to the typical Fano resonance and antiresonance structure 
in the transmission amplitude $T=|t|^2$ and $T=0$ at the antiresonance.
[See Fig.~\ref{ringdbrt}(a) and (f).]

%Class III
 When the magnetic AB flux is off an integer or a half-integer multiple of
the flux quantum, the imaginary part of the transmission zero is finite
and is sinusoidal as function of the AB phase $\phi$
[see the Eq.~(\ref{dotzero})].
That is, the zero of $t$ is shifted off the real-energy axis. 
When $0 < \phi < \pi$, the zero $Z_z$ lies 
in the upper half-plane of the complex-energy plane while 
the pole $Z_p$ lies in the lower half-plane. 
The trajectories of $t$ encircle the origin(the class III). 
The analytic expression of $\theta_t$ is 
\begin{eqnarray}
\theta_t 
 &=& \theta_f 
  + \tan^{-1} \frac{\epsilon - \epsilon_d + \overline{\Gamma} \sqrt{g/T_0} \cos\phi}
                   {\overline{\Gamma} \sqrt{g/T_0} \sin\phi}. 
\end{eqnarray}
The second term is the contribution from the zero. 
Since both the zero and the pole contribute the same sign of the 
phase by $\pi$ to $\theta_t$,  
$\theta_t$ evolves smoothly by the amount of $2\pi$ 
as shown in Fig.~\ref{ringdot}(b). 
The imaginary part of the zero is proportional to $\sin\phi$ 
and the zero moves away linearly with the magnetic field $B$ 
from the real-energy axis close to $\phi = n\pi$. 
The minimum value of $T$ (deriving from the transmission zero)
shows this trend as displayed in Fig.~\ref{ringdot}(c). 
As $\phi$ is increased from $0$ to $\pi/2$, the minimum value of $T_{\rm min}$ 
is increased and reaches the maximum at $\phi = \pi/2$. 
With further increase of $\phi$ from $\pi/2$ to $\pi$, 
$T_{\rm min}$ is reduced to zero.

%Class II
 When $\pi < \phi < 2\pi$, the pole and the zero 
lie in the same lower half-plane. 
The transmission amplitude, belonging to class II, delineates the 
closed orbit without encircling the origin. 
The transmission phase $\theta_t$ is given by the expression 
\begin{eqnarray}
\theta_t 
 &=& \theta_f + \pi 
  - \tan^{-1} \frac{\epsilon - \epsilon_d + \overline{\Gamma} \sqrt{g/T_0} \cos\phi}
                   {\overline{\Gamma} \sqrt{g/T_0} |\sin\phi|}. 
\end{eqnarray}
Since the pole and zero contribute the phase by $\pi$ but with the opposite sign
to $\theta_t$, the phase 
evolution is limited to the narrow range[see Fig.~\ref{ringdot}(e)] 
which is set by the two extreme 
points in the trajectory of $t$ viewed from the origin.

 In summary, we found three different classes for the transmission zeros. 
Though the behaviors of the phase evolution are different for the three classes, 
the transmission amplitudes remain in phase before and after the Fermi 
level is scanned through the real part of the transmission pole and zero. 
The transmission probability satisfies the Onsager relation, 
$|t_{RL}^{\phantom{*}}(-\phi)|^2 = |t_{RL}^{\phantom{*}}(\phi)|^2$
[see Fig.~\ref{ringdot}(c) and \ref{ringdot}(f)], 
even if the trajectories in the complex $t$ plot are different.
When the magnetic flux $\Phi$ is an 
integer or a half-integer multiples of the flux quantum,  
the transmission zero lies on the real-energy axis and 
the phase evolution of $\theta_t$ is featured with the abrupt jump 
by $\pi$ at the zero.
When the magnetic flux $\Phi$ is off from the 
integer values or the half-integer multiples of the flux quantum,  
the transmission zero lies off the real-energy axis and 
the phase evolution of $\theta_t$ becomes continuous over the energy scale 
$\overline{\Gamma} \sqrt{g} |\sin \phi|$ set by the magnetic fields.
Depending on the sign of the imaginary part of the transmission zero, 
the evolution of $\theta_t$ shows different behavior. 
When the zero lies in the upper half-plane, the contributions to the 
transmission phase from the zero-pole pair add up leading to 
the change of $2\pi$ over the zero-pole pair. 
On the other hand, two contributions are canceled by each other leading 
to the net change of $0$ in the transmission phase when 
the zero lies in the lower-half plane.

\section{\label{sectdbrt}AB ring with an embedded double-barrier well}
 In this section we study the scattering matrix of the AB ring when 
the double-barrier well (shown in Fig.~\ref{dbrtsystem}) 
is inserted in the lower arm.

 The symmetric double-barrier well can be described by the scattering matrix 
$S_0$ whose elements are given by the equations,
\begin{subequations}
\begin{eqnarray}
r_0 &=& r_0' ~=~ 
  \frac{\sqrt{1-T_0} (1-e^{2iKa}) } { 1 - (1-T_0)e^{2iKa} }, \\
t_0 &=& t_0' ~=~ 
  \frac{-T_0 e^{iKa} } { 1 - (1-T_0)e^{2iKa} }.
\end{eqnarray}
\end{subequations}
Here $K = \sqrt{k_F^2 + 2m eV_g/\hbar^2}$ is the wave number inside the 
double-barrier well, $a$ is the distance between two barriers, and 
$V_g$ is the gate voltage capacitatively coupled to the double-barrier well. 
The position of the resonant energy levels in the double-barrier well
is controlled by the gate voltage $V_g$. 
The incident electrons are confined to the Fermi level 
with the Fermi wave number $k_F$. 
Two barriers are assumed to be identical and to be described 
by the scattering matrix,
\begin{eqnarray}
\label{dbrttunsmtx}
S_b &=& \begin{pmatrix} \sqrt{1-T_0} & -i\sqrt{T_0} \cr
                 -i\sqrt{T_0} & \sqrt{1-T_0} \end{pmatrix}, 
\end{eqnarray}
where $T_0$ is the tunneling probability through the barrier. 
The transmission poles of $t_0$ (double-barrier well) are easily 
identified as
\begin{eqnarray}
K_pa &=& n\pi - i\Gamma, ~~\Gamma \equiv \frac{1}{2} \log\frac{1}{1-T_0}, 
\end{eqnarray}
where $n$ is a positive integer. If the barriers' scattering matrix 
is of the form given by the Eq.~(\ref{stubtunsmtx}) in Sec.~\ref{sectstub}, 
the poles are shifted by $\pi/2$.
% This difference comes from the different 
%boundary condition at the tunneling barriers. 
We note that $t_0$ can be expressed as the sum of simple poles,
\begin{eqnarray}
t_0 &=& - T_0 e^{iKa} \left[ \frac{1}{2}
    + \frac{i}{2}\sum_{n} \frac{1}{Ka - n\pi + i\Gamma} \right].
\end{eqnarray}
The transmission probability $|t_0|^2$ consists of  
a series of evenly spaced peaks with the same linewidth $\Gamma$. 
Transmission zeros are absent
in the double-barrier resonant tunneling system.

 We now study the properties of the transmission amplitude 
for the AB ring with an embedded double-barrier well. 
 Inserting the scattering matrix $S_0$ of the double-barrier well 
into the general expression of the $S$-matrix of the AB ring 
(Appendix \ref{app_abring}), 
we compute the transmission amplitude numerically. 
The results are displayed in Fig.~\ref{ringdbrt}. 
In the numerical works, we use the model parameters: 
$k_FL = 5\pi/3 (2\pi ~\mbox{mod}.)$;  
$\epsilon = 1/2, \lambda_1 = \lambda_2 = 1$ for the identical 
three-way splitters at the right and left junctions;
$T_0 = 0.2$ for the transmission probability of the 
double-barrier well.   
One closed orbit in the complex $t$ plot is completed with 
the variation of $\Delta (Ka) = 2\pi$. 
In the double-barrier well\cite{buttiker}, 
the orbit of $t$ is closed with the period $2\pi$ of $Ka$. 
Comparing our results to the Fig. 1 in the work\cite{buttiker}
 of Taniguchi and B\"{u}ttiker, the orbit of $t$ for the AB ring
is featured with an additional closed lobe. This lobe passes through 
the origin when $\phi = 0$ or $\pi$. 
But note that the lobe disappears in $t$ for some range of $\phi$. 
See the dotted line($\phi=135^{\circ}$) in Fig.~\ref{ringdbrt}(a) 
and the long dashed line($\phi=315^{\circ}$) in Fig.~\ref{ringdbrt}(d).

 When $\phi=0$ or $\pi$, the trajectory of $t$ passes through the origin
twice to complete the closed orbit. 
The transmission phase $\theta_t$ jumps by 
$\pi$ at the transmission zeros [see the solid line in Fig.~\ref{ringdbrt} (b)
and the dot-dashed line in Fig.~\ref{ringdbrt} (e)]
 since the transmission zeros are real.  
The phase increases by $\pi$ at one zero and decreases by $\pi$ 
at the other zero. 
These two real zeros are typical and behave differently 
when $\phi \neq 0$ or $\pi$, as will be shown later. 
Each transmission pole of the double-barrier well is paired with one
transmission zero in the AB ring. 
The transmission zeros in the AB ring are the consequence of  
the destructive interference between two arms. 
As shown in Fig.~\ref{ringdbrt}(c) and (f), 
the zero-pole pairs are developed in the order: zero-zero-pole-pole.

 When $\phi \neq 0$ or $\pi$, all trajectories of $t$ encircle 
the origin. These orbits can be considered as the combination of 
the two orbits: one(class III) encircles the origin
 while the other(class II) does not. 
Comparing $|t|^2$ and $\theta_t$ between Figs.~\ref{ringdot} and \ref{ringdbrt},
we can deduce that the transmission zeros are shifted off the 
real-energy axis. This point will be discussed later.

 Let us study the structure of the poles and the zeros in detail. 
We focus on the two consecutive zero-pole pairs: one near $Ka = 2\pi$
and the other near $Ka=3\pi$. These two zeros are typical
in the sense that others are the exact copies of these two 
in the physical properties.
The nature of these two zeros is different 
since they behave in the opposite way under the magnetic fields
(see Fig.~\ref{zerodbrt}.)
When $\phi=0$, we can deduce from Fig.~\ref{ringdbrt} (b) and (c) that 
the zero-pole pairs are ordered in the sequence: pole-zero-zero-pole. 
When $0 < \phi < \pi$, 
we conclude, using the results of the Sec.~\ref{sectdot},
for the zero-pole pair near $Ka=2\pi$ that 
the zero(pole) lies in the upper(lower) half-plane of the complex-energy plane,
respectively. 
On the other hand, the zero-pole pair near $Ka = 3\pi$ 
is in the lower half-plane. 
This is the reason why the phase change is $2\pi$ over the first zero-pole
pair and is $0$ over the second zero-pole pair. 
When $\phi=\pi$, the zero-pole pairs appear in the sequence: zero-pole-pole-zero.
When $\pi < \phi < 2\pi$, the roles of two afore-mentioned zero-pole pairs
are interchanged compared to the case of $0 < \phi < \pi$.

 We compute the traces of the transmission zeros $Z_z (\phi)$ 
in the complex-energy plane 
($z = Ka$) as function of the magnetic AB phase $\phi$. 
\begin{eqnarray}
Z_z (\phi) &=& E_z + \zeta(\phi). 
\end{eqnarray}
Here $E_z$ is the transmission zero when $\phi=0$ and $\zeta(\phi)$ 
is the shift of the zero in the presence of the magnetic AB flux. 
There are two distinct zeros in the AB ring with the double-barrier well
and the behavior of two zeros is different under the magnetic AB flux. 
The real and the imaginary parts of $\zeta(\phi)$ are plotted in Fig.~\ref{zerodbrt}
for two zeros of $E_z = Ka = 2\pi$ [panel (a)] and of 
$E_z = Ka \approx 2.8775\times\pi$ [panel (b)]. 
The real part of the first(second) zero is shifted downward(upward)
under the magnetic fields, respectively. 
The minimum in $T=|t|^2$ [see Fig.~\ref{ringdbrt} (c) and (e)] shows this trend.
%The sign of the imaginary part of the zeros cannot be deduced 
%from Figs.~\ref{ringdbrt}. 
The two zeros show the opposite behavior 
in their imaginary parts under the magnetic fields, too. 
The zero $E_z=2\pi (2.8775\times\pi)$ lies in the upper(lower) half-plane
when $0<\phi<\pi$, and in the lower(upper) half-plane when $\pi <\phi<2\pi$,
respectively. This corroborates our conclusion in the previous paragraph.
Two zeros are on the real-energy axis when $\phi = n\pi$ with $n$ being an integer.

 To summarize, there are two types of the transmission zeros which 
can be distinguished in their behavior under magnetic fields. 
Since one pole is always paired with one zero, the change in the 
transmission phase over the zero-pole pair is $0$ or $2\pi$
depending on the position of the zero in the complex-energy plane.

\section{\label{sectstub}AB interferometer with an embedded $T$-stub}
 In this section, we consider the AB ring with the side-branch or the $t$-stub.
The $t$-stub provides the different type of resonant levels
compared to the double-barrier well. 
In contrast to the double-barrier well, the $t$-stub 
itself provides the transmission zeros as well as the transmission poles. 
In this work, we consider the $t$-stub structure with two tunneling barriers 
which was previously studied in the literature\cite{sycho,lent}. 
We use the most symmetric 
three-way splitter at the junction with the parameters\cite{lent}, 
$\epsilon=4/9$, $\lambda_1 = -1$ and $\lambda_2 = 1$.
The double-barrier is assumed not to provide any resonant energy levels,
or the distance between two barriers is so short that the energy level
spacing is much larger than any other interesting energy scale in the problem. 
The introduction of two additional barriers to the $t$-stub
enables us to control the tunneling strength through the $t$-stub structure
and to mimic the quantum dot system.
The $S$-matrix of the two barriers is chosen to be
\begin{eqnarray}
\label{stubtunsmtx}
S_b &=& \begin{pmatrix} i\sqrt{1-T_0} & \sqrt{T_0} \cr
                 \sqrt{T_0} & i\sqrt{1-T_0} \end{pmatrix}, 
\end{eqnarray}
where $T_0$ is the tunneling probability through the barrier. 
We have chosen the different overall phase for $S_b$ compared to 
the double-barrier well in Sec.~\ref{sectdbrt}. 
The $S$-matrix of the $t$-stub with the double barriers is formulated 
in the Appendix \ref{app_dbtstub}.

 To get some insights on the position of the transmission zeros and poles, 
we consider the $S$-matrix of the simple $t$-stub. 
The incoming and outgoing current amplitudes can be matched at the junction 
of the $t$-stub. 
\begin{eqnarray}
\begin{pmatrix}O_S \cr O_L \cr O_R\end{pmatrix} 
 &=& \begin{pmatrix} \sigma_t & \sqrt{\epsilon_t} & \sqrt{\epsilon_t} \cr
              \sqrt{\epsilon_t} & a_t & b_t \cr 
              \sqrt{\epsilon_t} & b_t & a_t 
     \end{pmatrix} 
   \begin{pmatrix} I_S \cr I_L \cr I_R \end{pmatrix}.  
\end{eqnarray}
Here $I$'s and $O$'s are the incoming and outgoing current amplitudes 
at the junction. 
$\sigma_t = - a_t- b_t$ and $a_t,b_t$ can be determined to satisfy the unitarity 
of the scattering matrix. 
\begin{subequations}
\begin{eqnarray}
a_t &=& \frac{1}{2} \left[ \lambda_1 + \lambda_2 \sqrt{1 - 2\epsilon_t} \right], \\
b_t &=& \frac{1}{2} \left[ -\lambda_1 + \lambda_2 \sqrt{1 - 2\epsilon_t} \right].
\end{eqnarray}
\end{subequations}
There are four possible choices with $\lambda_i = \pm 1 (i=1,2)$. 
The value of $\epsilon_t$ is constrained: $0\leq \epsilon_t \leq 1/2$.  
When there is an infinite potential wall at the end of the stub of length $a$, 
$O_S$ and $I_S$ are related to each other by $I_S = O_S ~e^{i(2Ka + \pi)}$. 
The additional phase $\pi$ guarantees the node of the wave function 
at the infinite wall. 
The wave number $K$ in the stub is given by 
$K = \sqrt{k_F^2 + 2meV_g/\hbar^2}$. The quasibound state energy levels 
in the stub can be shifted with the gate voltage $V_g$ (capacitatively 
coupled to the $t$-stub). 
The effective $S$-matrix of the $t$-stub can be readily derived 
\begin{subequations}
\begin{eqnarray}
\begin{pmatrix} O_L \cr O_R \end{pmatrix} 
 &=& S_0 \begin{pmatrix}I_L \cr I_R\end{pmatrix}, ~~  
 S_0 ~=~ \begin{pmatrix}r_0 & t_0 \cr t_0 & r_0 \end{pmatrix},  \\
\label{zerostub1}
t_0 &=& b_t - \frac{\epsilon_t z}{1 + \sigma_t z } 
 ~=~ \frac{ b_t (1 + \lambda_1 z) }{1 - (a_t + b_t)z}, \\
r_0 &=& a_t - \frac{\epsilon_t z} {1 + \sigma_t z } 
 ~=~ \frac{ a_t (1 - \lambda_1 z) } { 1 - (a_t + b_t)z}. 
\end{eqnarray} 
\end{subequations}
Here $z=e^{2iKa}$. The first terms in $t_0$ and $r_0$ represent 
the direct scattering process 
and the second term comes from the multiple scattering processes in the stub. 
The unitarity of $S_0$ can be proved by showing that 
\begin{eqnarray}
&& |r_0|^2 + |t_0|^2 = 1, ~~~ t_0^*r_0 + r_0^* t_0 = 0
\end{eqnarray}
The transmission poles and zeros are located at 
\begin{subequations}
\begin{eqnarray}
\label{zerostub2}
Z_z &=& \left( n + \frac{1+\lambda_1}{4} \right) \pi, \\\
Z_p &=& \left( n + \frac{1-\lambda_2}{4} \right) \pi 
  - \frac{i}{2} \log \frac{1} {\sqrt{1-2\epsilon_t}}, 
\end{eqnarray} 
\end{subequations}
respectively. Here $n$ is an integer. For our choice of $\lambda_1 = -1$ 
and $\lambda_2 = 1$ for the $t$-stub, $Z_z = n\pi$ and $\mbox{Re}Z_p = n\pi$

 The scattering matrix of the $t$-stub with the double-barrier 
is derived in the Appendix~\ref{app_dbtstub}. 
Note that addition of the two barriers 
to the $t$-stub does not change the transmission zeros, but 
the poles are modulated by $T_0$ in both the real and the imaginary parts.

 The transmission amplitudes of the whole AB ring are computed 
numerically using the formulation detailed 
in the Appendices~\ref{app_abring} and \ref{app_dbtstub}
and are presented in Fig.~\ref{ringstub}.
The model parameters are: $\epsilon = 1/2$, $\lambda_1 = -1$, and $\lambda_2 = 1$
for the Shapiro matrices at the left and right three-way junctions.
The three-way splitter for the $t$-stub was chosen to be the most symmetric 
one as noted above. The tunneling barriers for the $t$-stub are chosen 
as $T_0 = 0.8$.

As can be deduced from the Fig.~\ref{ringstub} (c) and (f), the zero-pole
pairs appear in the order: zero-pole-zero-pole. 
All the trajectories of $t$ are circles and the closed orbit is completed 
with $\Delta Ka = \pi$. Three different classes of orbits of $t$ are 
realized for this system varying the magnetic AB flux. 
%In contrast with the DBRT system, the $t$-stub itself provides 
%the transmission zeros as noted above. 

 When $\phi=0$ or $\phi=\pi$, the transmission zero lies on the 
real-energy axis and the transmission phase jumps by $\pi$ at the 
zeros. When $\phi=0$, $\theta_t$ near $Ka=2\pi$ drops by $\pi$ at the zero 
and increases smoothly by the amount $\pi$ due to the pole. 
For this zero-pole pair, the zero precedes the pole as shown
in Fig.~\ref{ringstub} (b) and (c).  
When $\phi=\pi$, $\theta_t$ drops by $\pi$ at the zero and increases 
almost linearly due to the pole. We can deduce from the functional shape 
of the $\theta_t$ and $T=|t|^2$ [see Fig.~\ref{ringstub} (e) and (f)]
that the zeros and poles are almost evenly interlaced.

 When $0 < \phi < \pi$, the orbits of $t$ encircle the origin
and the phase evolution of $\theta_t$ is smooth and continuous. 
The orbits move away from the origin with increasing the AB phase. 
This trend is clearly visible in Fig~\ref{ringstub} (a) and (c). 
Since the zeros lie in the upper half-plane and the poles 
are in the lower half-plane, two contributions to $\theta_t$ add up 
and lead to the change of $\theta_t$ by $2\pi$ over the zero-pole pair.

 The orbits of $t$ lie outside the origin when $\pi < \phi < 2\pi$. 
In this case, the zeros and the poles lie in the same lower half-plane.
Their contributions to the transmission phase are opposite in sign, 
and the net change of $\theta_t$ over the zero-pole
pair is zero. Since the phase decrease precedes the phase increase,
the zero precedes the pole. The zero-pole pair approaches each other
as the value of $\phi$ is increased from $\pi$ to $2\pi$.

 The trace of one typical transmission zero $Z_z (\phi)$ 
in the complex-energy plane ($z = Ka$) is computed and 
plotted in Fig.~\ref{zerostub} 
as function of the magnetic AB phase $\phi$. 
\begin{eqnarray}
Z_z (\phi) &=& E_z + \zeta(\phi). 
\end{eqnarray}
Here $E_z$ is the transmission zero when $\phi=0$ and $\zeta(\phi)$ 
is the shift of the zero in the presence of the magnetic AB flux. 
In contrast to the AB ring with the double-barrier well, 
the nature of all zeros is identical in the sense that their behavior is
the same under the magnetic fields. 
The real and imaginary parts of $\zeta(\phi)$ are plotted in Fig.~\ref{zerostub}
for the zero of $E_z = Ka \approx 1.9565\times\pi$.  
Note that this zero in the AB ring is shifted from $E_z = 2\pi$ of the $t$-stub with 
the double-barrier. 
As expected from the shift of the minimum position of $T=|t|^2$ 
[see Fig.~\ref{ringstub} (c) and (e)], the real part of the zero is
positively shifted in the presence of the magnetic fields. 
The imaginary part of the zero is sinusoidal as function of the AB phase
$\phi$ and vanishes when the magnetic AB flux is an integer or a half-integer
multiple of the flux quantum.

\section{\label{sectsum}Summary and Conclusion}
 In this paper, we studied the behavior of the transmission zeros 
and the corresponding changes in the transmission phase when 
the time-reversal symmetry of the system is broken by 
magnetic fields. For our study we considered 
the Aharonov-Bohm(AB) interferometers with one scattering center in 
the lower arm. Studied scattering centers include 
the system of one discrete energy level, the double-barrier
well, and the $t$-stub with the double barriers. 
Each resonant level in the scattering center gives rise to 
a transmission pole and is paired with a
transmission zero in the AB ring. 
Due to the causality relation, 
the transmission pole always lies in the lower half-plane 
of the complex-energy plane. 
On the other hand, the zero can be anywhere in the complex-energy plane
and its position can be controlled by the magnetic AB flux.
Depending on the position of the transmission zeros in the complex-energy 
plane, the trajectory and the phase of the transmission amplitude 
show different behaviors.

 The transmission zeros lie on the real-energy
axis when the magnetic AB flux is an integer or a half-integer 
multiple of the flux quantum. The transmission phase jumps by $\pi$
at the transmission zeros in this case.

The transmission zeros are shifted off the real-energy axis
and can be either in the upper or in the lower half-plane of the 
complex-energy plane,    
when the AB magnetic flux is off from the integer or the half-integer 
multiples of the flux quantum.
The evolution of the transmission phase in this case 
is continuous as the Fermi level is scanned through 
the real part of the transmission zeros.

When the zeros lie in the lower half-plane, the orbits of 
the transmission amplitude lie outside the origin 
and the phase change is limited by the two extreme points 
of the orbit when viewed from the origin. 
Since the zero-pole pair contributes the opposite sign 
of the phase by $\pi$, the net change in the 
transmission phase over the zero-pole is zero.

The orbits of the transmission amplitude encircle 
the origin when the zeros lie in the upper-half plane.
In this case, the zero-pole pair give the same sign of 
the phase by $\pi$ to the transmission phase, 
the total accrued phase over this zero-pole pair is $2\pi$.

 Though the zero-pole pair leads to the different phase 
evolution depending on the position of the zero in the 
complex-energy plane, the transmission phase 
remains in phase after passing through the zero-pole pair.

 The modulation of the transmission zeros in the closed AB ring 
may be tested in experiments using the following setup.
 We may insert the closed AB ring 
into  one arm of the larger AB ring. 
The larger AB ring should be the open system where the multiple 
windings of the electrons are prevented. 
The evolution of the transmission phase in the closed AB ring
can be measured by making the period of the AB oscillation in the larger ring 
much shorter than that of the closed AB ring.

 Entin-Wohlman {\it et. al.}\cite{israel} studied the effects of broken 
unitarity on the phase locking. According to their study, the phase jump
at the Coulomb peaks can become smooth by breaking the unitarity 
of the AB ring. In our work, the phase jump at the transmission zeros
is shown to become continuous by breaking the time-reversal symmetry. 
Recently, Kobayashi {\it et. al.}\cite{japan} studied experimentally the tuning
of the Fano effect in the AB ring with an embedded quantum dot.   
Some of their results might be relevant to our theoretical results.

\acknowledgments
 We are indebted to H. W. Lee for useful discussions. 
This work was supported in part by the BK21 project and
in part by grant No. 1999-2-114-005-5 from the KOSEF 
and by the Center for Strongly Correlated 
Materials Research(SNU) from the KOSEF.

\appendix

\section{\label{app_abring}$S$-matrix of Aharonov-Bohm ring}
 In this Appendix, we derive the $S$-matrix of an AB ring in a compact form 
when some scattering centers are present on the AB ring 
and especially when the interesting
target system is inserted in the lower arm of the AB ring as shown 
in Fig.~\ref{abring}.  
The amplitudes of the incoming and outgoing 
waves at the left and right junctions 
are related to each other by the scattering matrix.
\begin{subequations}
\begin{eqnarray}
\begin{pmatrix} O_L \cr x_1 \cr x_2 \end{pmatrix} 
 &=& S^L \begin{pmatrix}I_L \cr y_1 \cr y_2 \end{pmatrix}, 
\begin{pmatrix}O_R \cr v_1 \cr v_2\end{pmatrix} 
 = S^R \begin{pmatrix}I_R \cr u_1 \cr u_2\end{pmatrix}, \\
S^{p} &=& \begin{pmatrix}\sigma_p & \sqrt{\epsilon_p} & \sqrt{\epsilon_p} \cr 
       \sqrt{\epsilon_p} & a_p  & b_p \cr
       \sqrt{\epsilon_p} & b_p & a_p \end{pmatrix}, ~~p=L,R.
\end{eqnarray}
\end{subequations}
Here $S^{L,R}$ are the Shapiro matrices responsible for the splitting of the 
electron wave functions in three pathways. 
The unitarity leads to four possible solutions, 
\begin{subequations}
\begin{eqnarray}
\sigma_p &=& - a_p - b_p, \\
a_p &=& \frac{1}{2} \left[ 
   \lambda^p_1 + \lambda^{p}_{2} \sqrt{1 - 2\epsilon_p} \right], \\
b_p &=& \frac{1}{2} \left[ 
   -\lambda^p_1 + \lambda^{p}_{2} \sqrt{1 - 2\epsilon_p} \right], 
\end{eqnarray}
\end{subequations}
where $\lambda^{p}_1, \lambda^{p}_2 = \pm 1$. 
To simplify the algebra, we introduce new notations, 
\begin{eqnarray}
S_{p} &\equiv& \begin{pmatrix} a_p & b_p \cr b_p & a_p\end{pmatrix}, ~~
|s_p> ~\equiv~ \begin{pmatrix} \sqrt{\epsilon_p} \cr \sqrt{\epsilon_p} \end{pmatrix}. 
\end{eqnarray}
$S_{L,R}$ is the $2\times 2$ submatrix of $S^{L,R}$, respectively. 
The amplitudes of waves 
at the left and right junctions are related to each other by the 
scattering matrix which is responsible for the scattering processes 
in the two arms, 
\begin{eqnarray}
\label{eqn1}
\begin{pmatrix} |y> \cr |u> \end{pmatrix}
 &=& \begin{pmatrix} {\cal R} & {\cal T}' \cr 
               {\cal T} & {\cal R}' \end{pmatrix}
   \begin{pmatrix} |x> \cr |v>\end{pmatrix}.
\end{eqnarray}
The ket vectors are defined, e.g., as 
$|x > \equiv \left(\begin{smallmatrix}x_1 \cr x_2\end{smallmatrix}\right)$. 
${\cal R}$ and ${\cal T}$ are the $2\times 2$ matrices 
which contain the information of 
the scattering matrices in each arm and the phase accrued by the motion of
electrons along the ring. 
These matrices are model-specific and are discussed in the main text. 
We want to find the $S$-matrix of the ring. 
\begin{subequations}
\begin{eqnarray}
\label{eqn2}
O_L &=& \sigma_L I_L + <s_L | y>, \\
O_R &=& \sigma_R I_R + <s_R | u>, \\
\label{eqn3}
|x> &=& I_L |s_L> + S_L |y>, \\
|v> &=& I_R |s_R> + S_R |u>. 
\end{eqnarray}
\end{subequations}
From the above equations, it is straightforward 
to derive the following results,
\begin{subequations}
\begin{eqnarray}
|y> &=& I_L \cdot [ 1 - \overline{\cal R} S_L ]^{-1} \overline{\cal R} |s_L> 
     \nonumber\\
   && + I_R \cdot [ 1 - \overline{\cal R} S_L ]^{-1} \overline{\cal T}' |s_R>,  \\
|u> &=& I_L \cdot [ 1 - \overline{\cal R}' S_R ]^{-1} \overline{\cal T} |s_L> 
     \nonumber\\
   && + I_R \cdot [ 1 - \overline{\cal R}' S_R ]^{-1} \overline{\cal R}' |s_R>, 
\end{eqnarray}
\end{subequations}
where newly defined reflection and transmission matrices are given by the 
expressions, 
\begin{subequations}
\begin{eqnarray}
\overline{\cal R}
 &=& {\cal R} + {\cal T}' [1 - S_R {\cal R}']^{-1} S_R {\cal T},   \\
\overline{\cal T} 
 &=& {\cal T} [1 - S_L {\cal R}]^{-1},   \\
\overline{\cal R}'
 &=& {\cal R}' + {\cal T} [1 - S_L {\cal R}]^{-1} S_L {\cal T}',  \\
\overline{\cal T}' 
 &=& {\cal T}' [1 - S_R {\cal R}']^{-1}. 
\end{eqnarray}
\end{subequations}
After more algebra, we find the $S$-matrix of the ring,
\begin{subequations}
\begin{eqnarray}
\begin{pmatrix}O_L \cr O_R\end{pmatrix}
 &=& S_{\rm ring} \begin{pmatrix}I_L \cr I_R\end{pmatrix},  \\
S_{\rm ring} 
 &=& \begin{pmatrix} r_{LL}^{\phantom{*}} & t_{LR}^{\phantom{*}} \cr 
     t_{RL}^{\phantom{*}} & r_{RR}^{\phantom{*}} \end{pmatrix}, \\
r_{LL}^{\phantom{*}}
 &=& \sigma_L + <s_L| [ 1 - \overline{\cal R} S_L ]^{-1} \overline{\cal R} |s_L>, \\
r_{RR}^{\phantom{*}}
 &=& \sigma_R + <s_R| [ 1 - \overline{\cal R}' S_R ]^{-1} \overline{\cal R}' |s_R>, \\
t_{LR}^{\phantom{*}}
 &=& <s_L| [ 1 - \overline{\cal R} S_L ]^{-1} \overline{\cal T}' |s_R>, \\
t_{RL}^{\phantom{*}}
 &=& <s_R| [ 1 - \overline{\cal R}' S_R ]^{-1} \overline{\cal T} |s_L>. 
\end{eqnarray}
\end{subequations}

\section{\label{app_dbtstub}Scattering matrix of the $t$-stub with double barriers}
 We consider the $S$-matrix of the $t$-stub with the double barriers 
[see Fig.~\ref{stubsystem}]. 
The distance between two barriers is assumed to be too short to allow the 
resonant energy levels, or the energy level spacing is very large compared to 
other energy scales. 
But the length of the stub, $a$, is long enough to allow the quantized 
energy levels in the isolated stub. 
In this case, the amplitude of the wave functions can 
be matched as
\begin{subequations}
\begin{eqnarray}
\begin{pmatrix} O_S \cr x_1 \cr x_2 \end{pmatrix}
 &=& S^{t} \begin{pmatrix}I_S \cr y_1 \cr y_2 \end{pmatrix}, ~
 S^{t} ~=~ \begin{pmatrix} \sigma & \sqrt{\epsilon} & \sqrt{\epsilon} \cr
              \sqrt{\epsilon} & a & b \cr 
              \sqrt{\epsilon} & b & a \end{pmatrix}, \\
\begin{pmatrix}O_L \cr y_1\end{pmatrix} 
 &=& S_L \begin{pmatrix}I_L \cr x_1\end{pmatrix}, ~  
 \begin{pmatrix}O_R \cr y_2\end{pmatrix} 
  ~=~ S_R \begin{pmatrix}I_R \cr x_2\end{pmatrix},  \nonumber\\
 && S_p ~=~ \begin{pmatrix}r_p & t_p' \cr t_p & r_p'\end{pmatrix}, ~~p=L,R. 
\end{eqnarray}
\end{subequations}
We want to find the $S$-matrix of the $t$-stub with the double barriers,
\begin{eqnarray}
\begin{pmatrix} O_L \cr O_R\end{pmatrix} 
 &=& S \begin{pmatrix}I_L \cr I_R\end{pmatrix}.
\end{eqnarray} 
When there is an infinite potential wall at the end of the stub, two amplitudes
$I_s$ and $O_s$ are constrained as $I_s = e^{i(2Ka+\pi)} O_s$.
We can rewrite the above relations between the amplitudes as
\begin{subequations}
\begin{eqnarray}
O_s &=& \sigma I_s + <s_t |y>, \\
|x> &=& I_s |s_t> + S_t |y>, \\
|O> &=& R |I> + T' |x>, \\
|y> &=& T |I> + R' |x>.
\end{eqnarray} 
\end{subequations}
New notations are introduced to simplify the algebra.
\begin{subequations}
\begin{eqnarray}
|x> &=& \begin{pmatrix}x_1 \cr x_2\end{pmatrix}, ~~
 |y> ~=~ \begin{pmatrix}y_1 \cr y_2\end{pmatrix}, \\
|I> &=& \begin{pmatrix}I_L \cr I_R\end{pmatrix}, ~~ 
 |O> ~=~ \begin{pmatrix}O_L \cr O_R\end{pmatrix}, \\
|s_t> &=& \begin{pmatrix}\sqrt{\epsilon} \cr \sqrt{\epsilon} \end{pmatrix}, ~~
 S_t ~=~ \begin{pmatrix} a & b \cr b & a\end{pmatrix}, \\
R &=& \begin{pmatrix} r_L & 0 \cr 0 & r_R\end{pmatrix}, ~~
 T ~=~ \begin{pmatrix} t_L & 0 \cr 0 & t_R\end{pmatrix}, \nonumber\\
R' &=& \begin{pmatrix} r_L' & 0 \cr 0 & r_R'\end{pmatrix}, ~~
 T' ~=~ \begin{pmatrix} t_L' & 0 \cr 0 & t_R'\end{pmatrix}.
\end{eqnarray}
\end{subequations}
From the relation between $I_s$ and $O_s$, we find
\begin{eqnarray}
I_s &=& - \frac{1} {\sigma + e^{-2iKa} } ~ <s_t |y> \nonumber\\
 &=& - \frac{ <s_t| [1 - R'S_t]^{-1} T |I> } 
   { \sigma + e^{-2iKa} + <s_t|  [1 - R'S_t]^{-1} R' |s_t> }.
\end{eqnarray}
After some algebra, we find the $S$-matrix of the system $|O> = S_0 |I>$,
\begin{eqnarray}
S_0 &=& R + T' [1 - S_t R']^{-1} S_t T  \nonumber\\
 && - \frac{ T'[1 - S_t R']^{-1} |s_t> <s_t| [1 - R'S_t]^{-1} T } 
    { \sigma + e^{-2iKa} + <s_t|  [1 - R'S_t]^{-1} R' |s_t> }.
\end{eqnarray}
Note that the transmission poles are determined by the zeros of 
$\sigma + e^{-2iKa} + <s_t|  [1 - R'S_t]^{-1} R' |s_t>$.

 For the two identical barriers described by the scattering matrix 
($R_0 = 1 - T_0$),
\begin{eqnarray}
S_b &=& \begin{pmatrix} i\sqrt{R_0} & \sqrt{T_0} \cr 
            \sqrt{T_0} & i\sqrt{R_0} \end{pmatrix},
\end{eqnarray}
the scattering matrix of the $t$ stub with double-barrier
is given by the equations, 
\begin{subequations}
\begin{eqnarray}
S_0 &=& \begin{pmatrix} r_0 & t_0 \cr t_0 & r_0 \end{pmatrix}, \\
t_0 &=& \frac{T_0 b}{(1-i\sqrt{R_0}a)^2 + R_0 b^2} 
   \frac{e^{-2iKa} + \lambda_1} 
    { e^{-2iKa} + \frac{\sigma + i\sqrt{R_0}}{1+i\sqrt{R_0} \sigma} }, \\
r_0 &=& i\sqrt{R_0} + \frac{T_0}{1+i\sqrt{R_0} \sigma} \cdot
  \left[ \frac{a + i\sqrt{R_0}(b^2-a^2)} {1 - i\sqrt{R_0}(a-b)} \right. \nonumber\\
 && \left. - \frac{\epsilon} 
      { (1+i\sqrt{R_0} \sigma)e^{-2iKa} + \sigma + i\sqrt{R_0} }
   \right]. 
\end{eqnarray}
\end{subequations}
Note that the transmission zeros do not depend on or are not modified by
the barriers' tunneling strength $T_0$ [see Eqs.~(\ref{zerostub1}) and (\ref{zerostub2})], 
but the poles are modulated by the value of $T_0$.
The transmission poles($Z_p$) and zeros($Z_z$) are located at
\begin{subequations} 
\begin{eqnarray}
Z_z &=& \left( n + \frac{1+\lambda_1}{4} \right) \pi, \\\
Z_p &=& \left( n + \frac{1-\lambda_2}{4} \right) \pi  \nonumber\\
 && + \frac{\lambda_2}{2} \left[ \tan^{-1} \sqrt{\frac{R_0}{1-2\epsilon}} 
   - \tan^{-1} \sqrt{R_0(1-2\epsilon)} \right] \nonumber\\
 && - \frac{i}{2} \log \sqrt{ \frac{1+R_0(1-2\epsilon)} {1-2\epsilon + R_0} }. 
\end{eqnarray} 
\end{subequations}
As expected, the linewidth 
of quasibound states in the stub (or the imaginary part of poles)
is reduced with the reduced $T_0$.

\begin{widetext}

\newpage

%%%%
%%%%
%%%%  All figures are collected here
%%%%
%%%%

%
%
\begin{figure}
\includegraphics{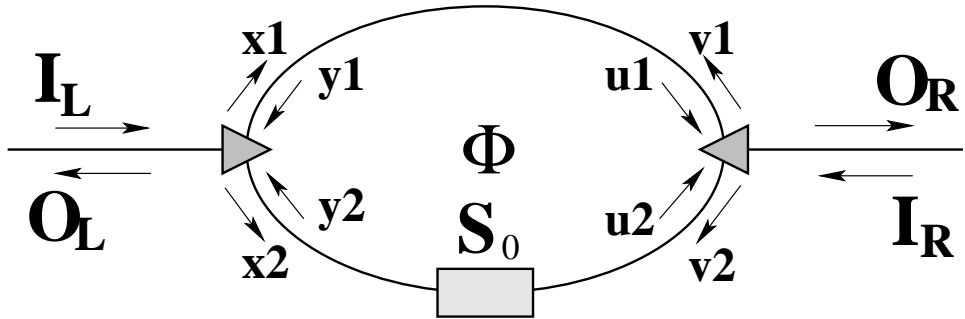} \vspace{0.5cm} 
\caption{Aharonov-Bohm ring with an embedded target system. 
The scattering process in the target system is described by the 
scattering materix $S_0$. The length of the upper and lower arms
is denoted by $L$. $\Phi$ is the magnetic Aharonov-Bohm(AB) flux
threading through the AB ring. \label{abring}}
\end{figure}
\begin{figure}
\includegraphics{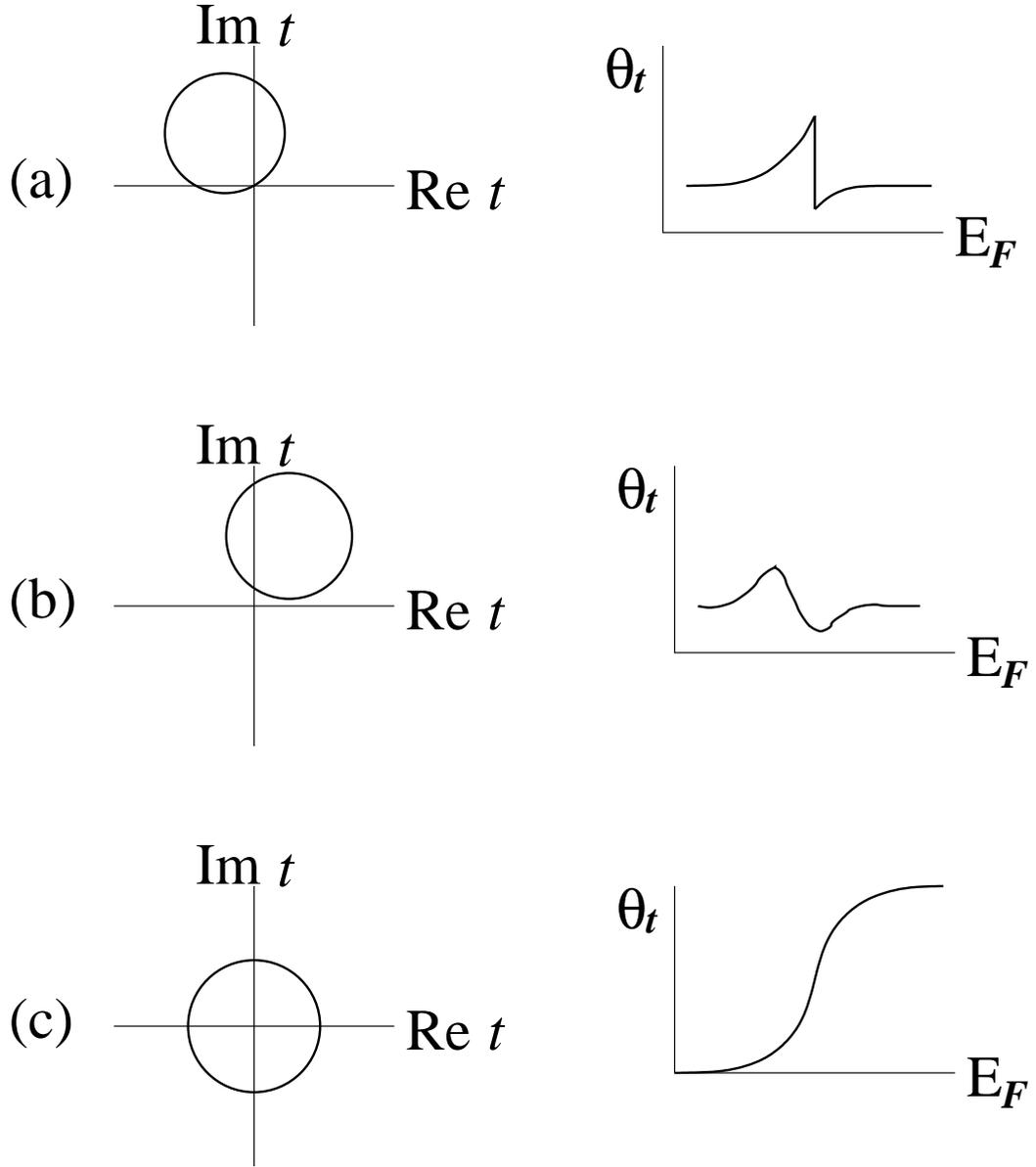}
\vskip 0.5cm
\caption{Schematic display of trajectories of the transmission amplitude $t$ 
and its phase variation. In the panel (a), the imaginary part of the
transmission zero $Z_z$ is null or \mbox{Im}$Z_z = 0$ (Class I). 
In (b), \mbox{Im} $Z_z < 0$ (Class II) and in (c), \mbox{Im} $Z_z > 0$ (Class III). 
\label{3tzero}}
\end{figure}
\begin{figure}
\includegraphics{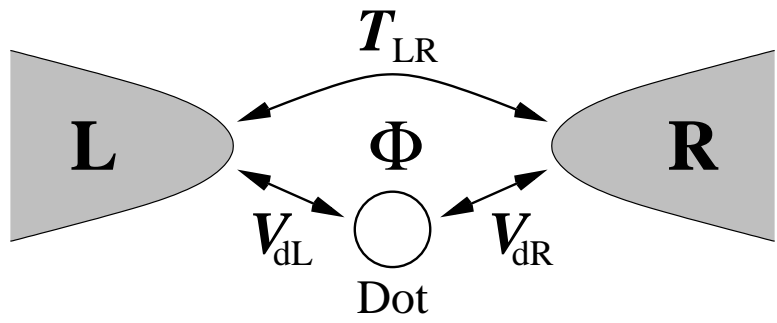}
\vskip 0.5cm
\caption{Schematic display of Aharonov-Bohm(AB) interferometer with a quantum dot.
The dot is modeled by one discrete energy level.
\label{dotsystem}}
\end{figure}
\begin{figure}
\noindent
\begin{minipage}[t]{0.45\linewidth}
 \includegraphics{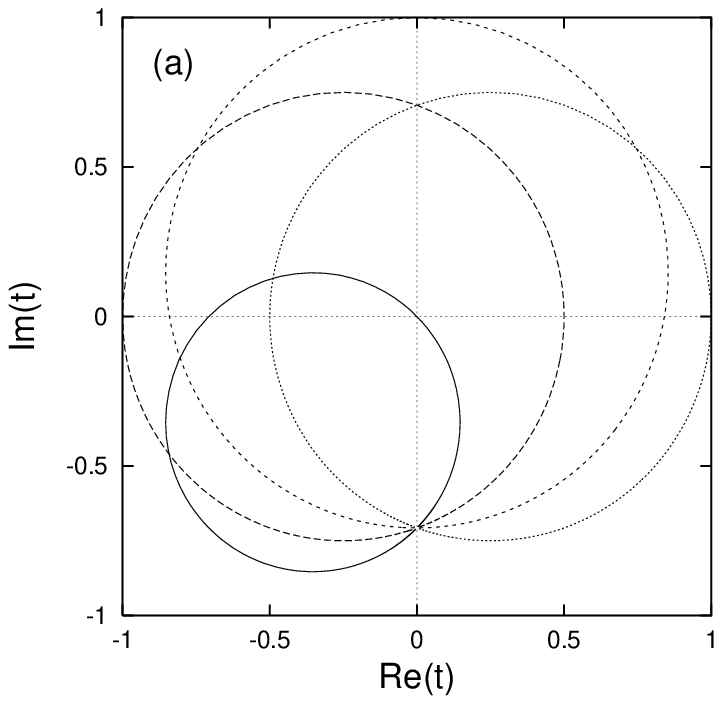}
 \includegraphics{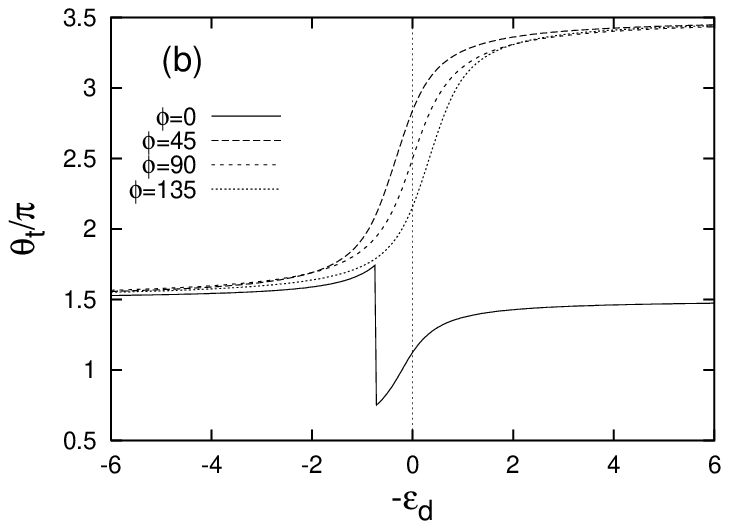}
 \includegraphics{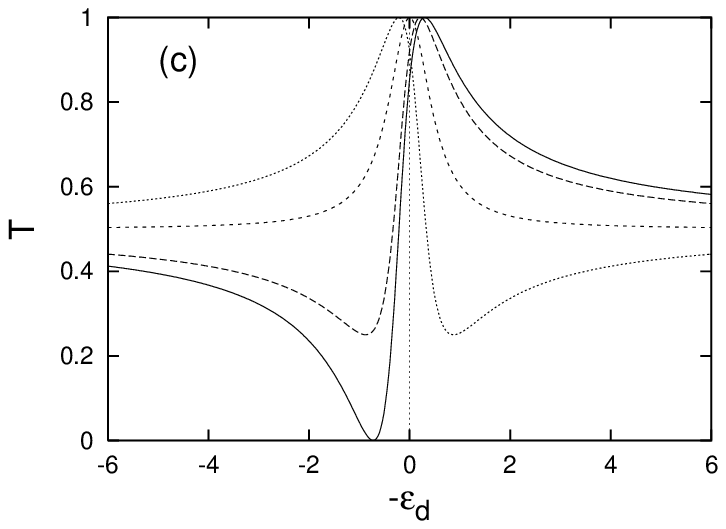}
\end{minipage} 
\begin{minipage}[t]{0.45\linewidth}
 \includegraphics{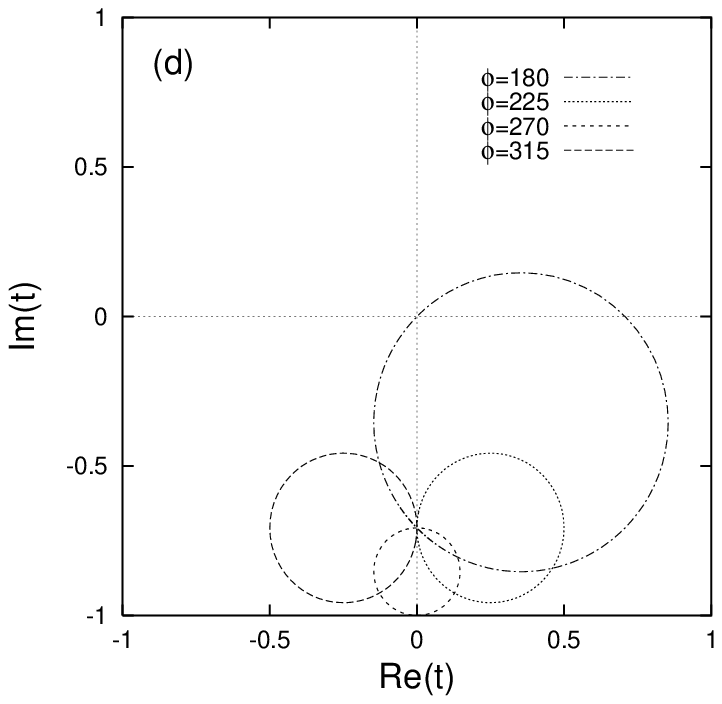}
 \includegraphics{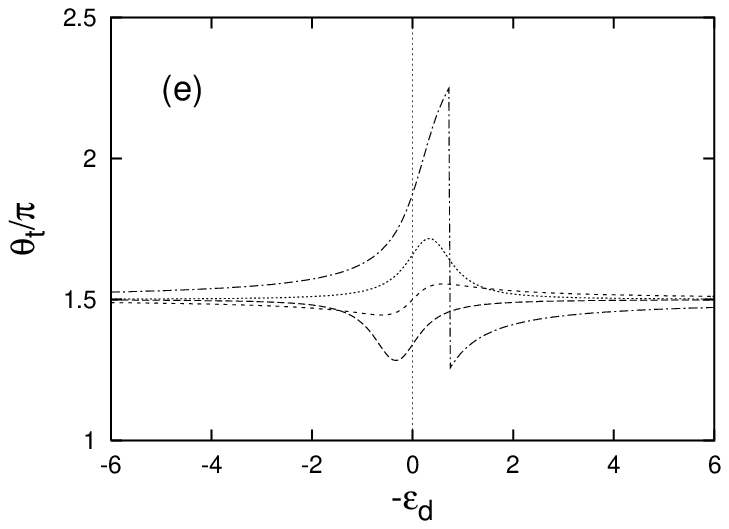}
 \includegraphics{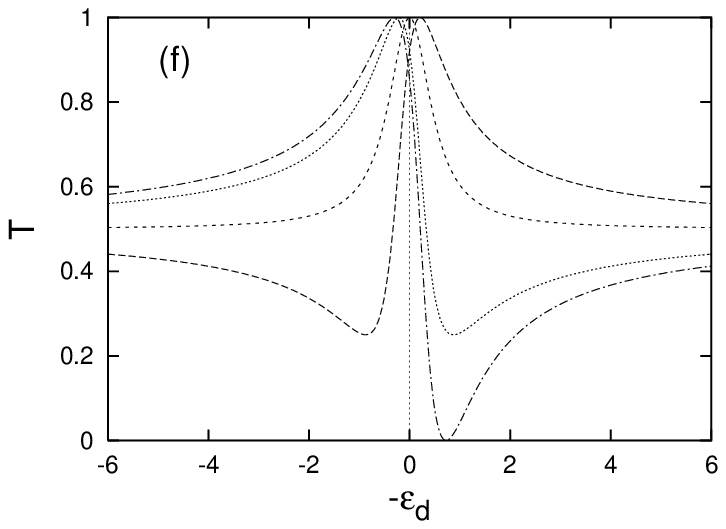}
\end{minipage} 
\vskip 0.5cm 
\caption{Behavior of the transmission amplitude $t$ with varying the AB phase $\phi$
for the AB interferometer with one discrete energy level.
Panels (a) and (d) display the trajectories of $t$ as an implicit function 
of the discrete energy level $\epsilon_d$. The evolution of the 
transmission phase $\theta_t$ is shown in panels (b) and (e). 
The transmission probability $T=|t|^2$ is displayed in two panels (c) and (f). 
The AB phases are the same for the same lines 
either in the left column panels (a), (b) and (c) 
or in the right column panels (d), (e) and (f). 
\label{ringdot}}
\end{figure}
\begin{figure}
\includegraphics{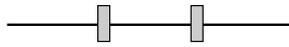}
\vskip 0.5cm
\caption{Double-barrier well. \label{dbrtsystem}}
\end{figure}
\begin{figure}
\noindent
\begin{minipage}[t]{0.45\linewidth}
 \includegraphics{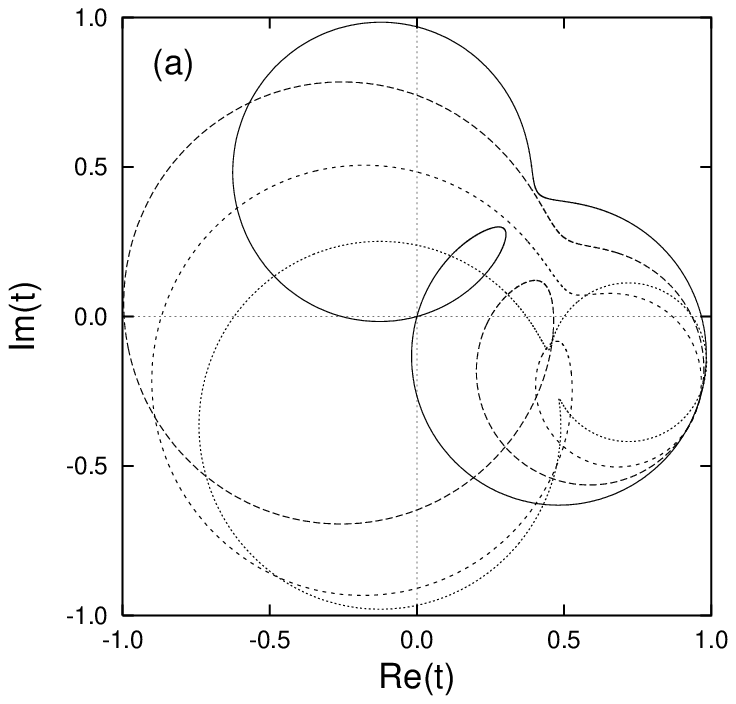}
 \includegraphics{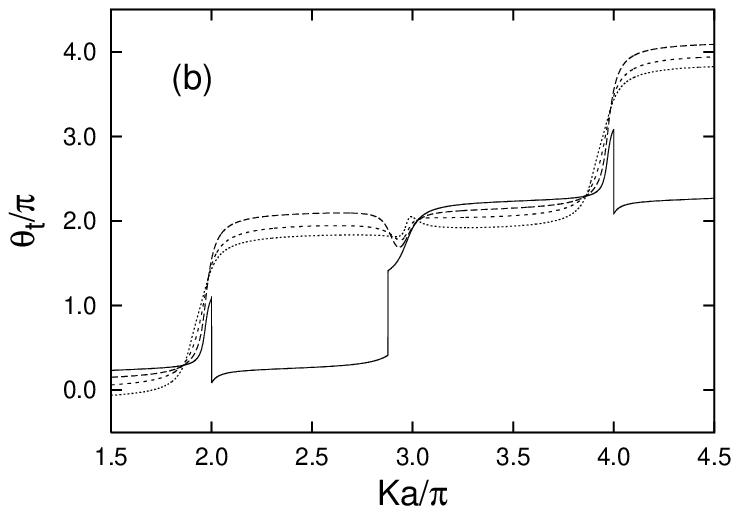}
 \includegraphics{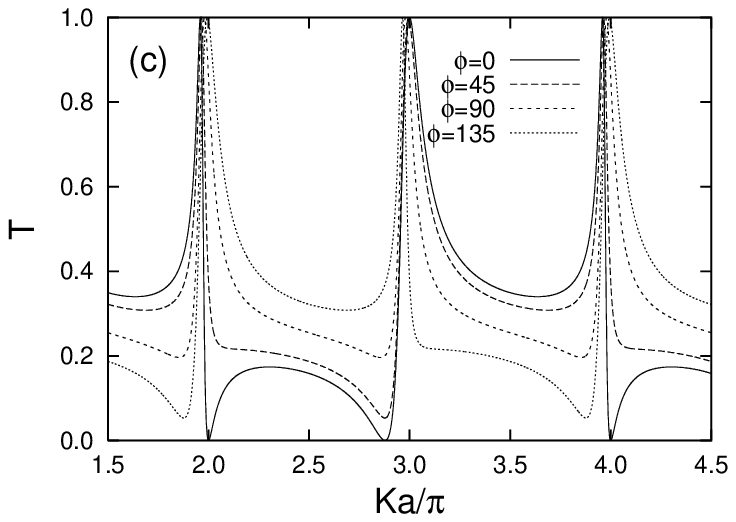}
\end{minipage} 
\begin{minipage}[t]{0.45\linewidth}
 \includegraphics{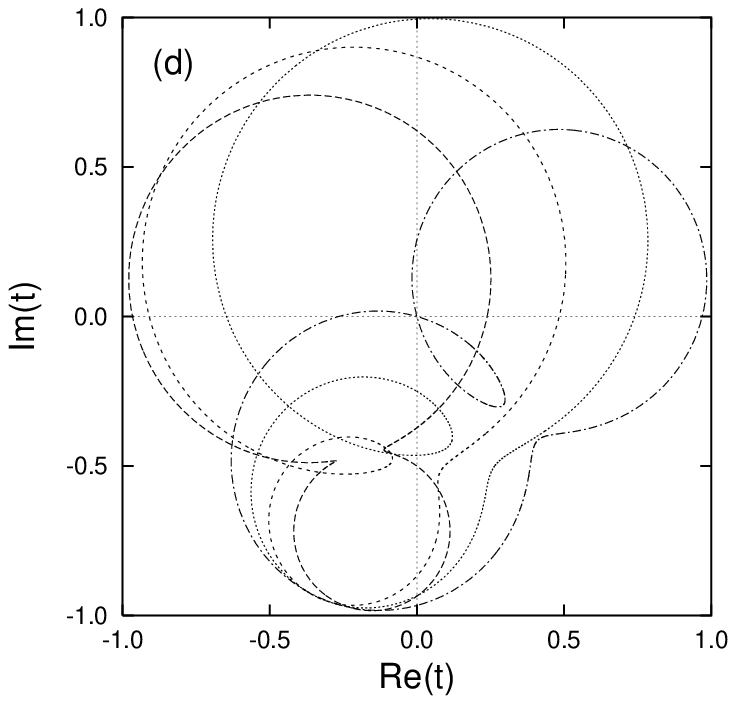}
 \includegraphics{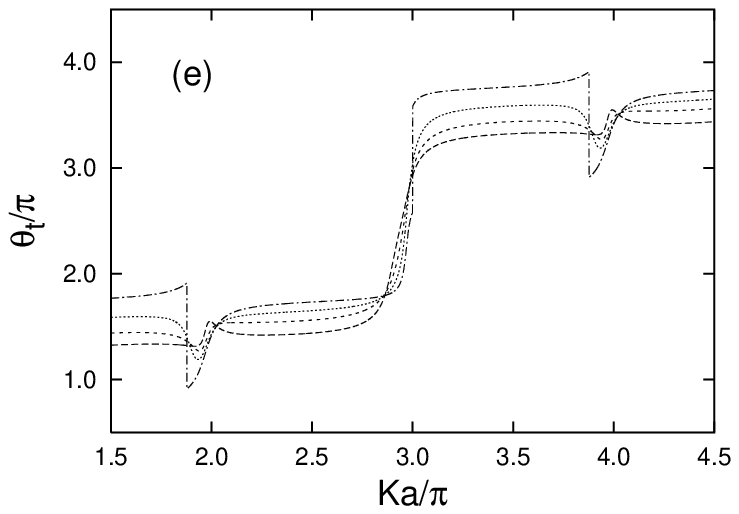}
 \includegraphics{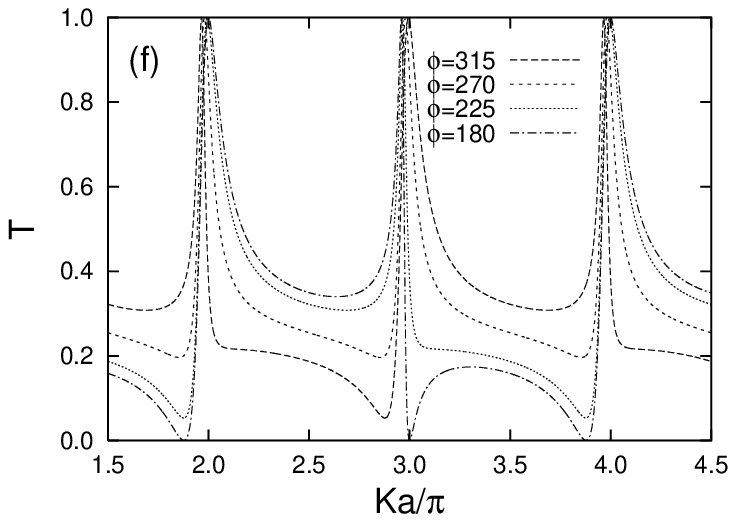}
\end{minipage} 
\vskip 0.5cm 
\caption{Behavior of the transmission amplitude $t$ with varying the AB phase $\phi$
for the AB ring with the double-barrier well. 
Panel descriptions are the same as in Fig.~\ref{ringdot}.
Model parameters are chosen as 
$k_FL = 5\pi/3 (\mbox{mod.}~ 2\pi)$, $\epsilon_{L,R} = 1/2$, 
$\lambda_1 = \lambda_2 = 1$ and $T_0 = 0.2$.
\label{ringdbrt}}
\end{figure}
\begin{figure}
\includegraphics{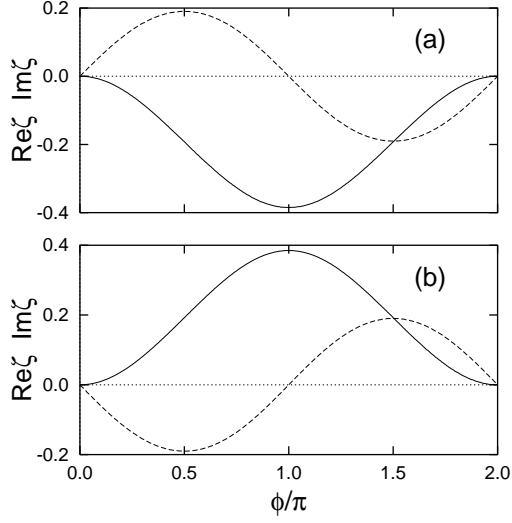}
\vskip 0.5cm
\caption{Transmission zeros as function of the AB phase $\phi$
in the AB ring with the double-barrier well.  
The AB phase dependence of the zero at $Ka=2\pi [2.8775\times\pi]$ is displayed
 in the panel (a)[(b)], respectively. 
Solid(dashed) line is the real(imaginary) 
part of the shifted zero $\zeta(\phi)$, respectively. 
The transmission zero is represented by $Z_z (\phi) = E_z + \zeta(\phi)$ 
where $E_z$ is the transmission zero when $\phi=0$ and $\zeta(\phi)$ 
is the shift of the zero in the presence of the magnetic AB flux.    
\label{zerodbrt}}
\end{figure}
\begin{figure}
\includegraphics{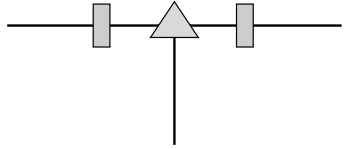}
\vskip 0.5cm
\caption{$t$-stub with the double-barrier.
\label{stubsystem}}
\end{figure}
\begin{figure}
\noindent
\begin{minipage}[t]{0.45\linewidth}
 \includegraphics{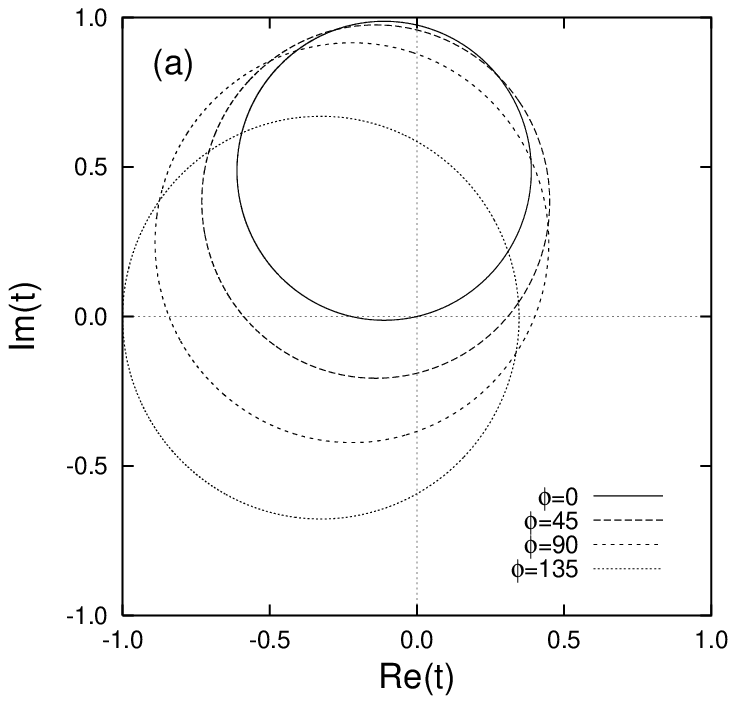}
 \includegraphics{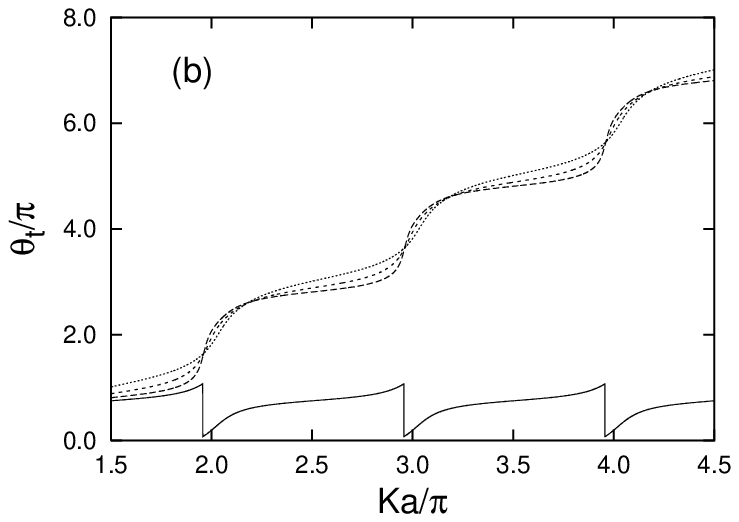}
 \includegraphics{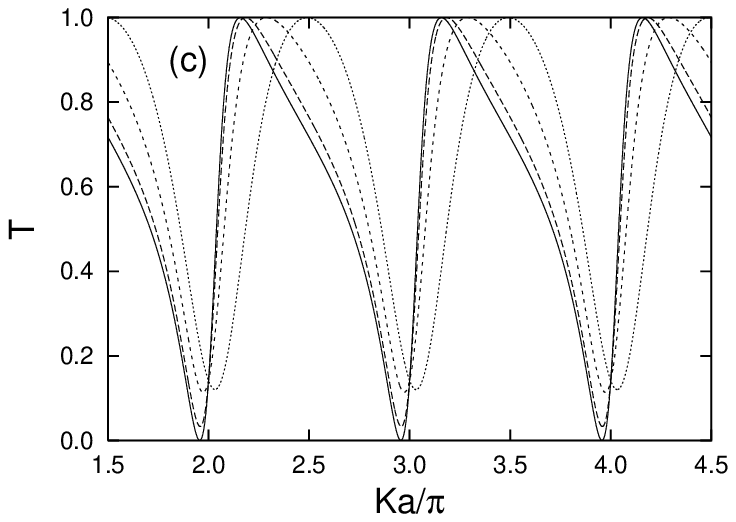}
\end{minipage} 
\begin{minipage}[t]{0.45\linewidth}
 \includegraphics{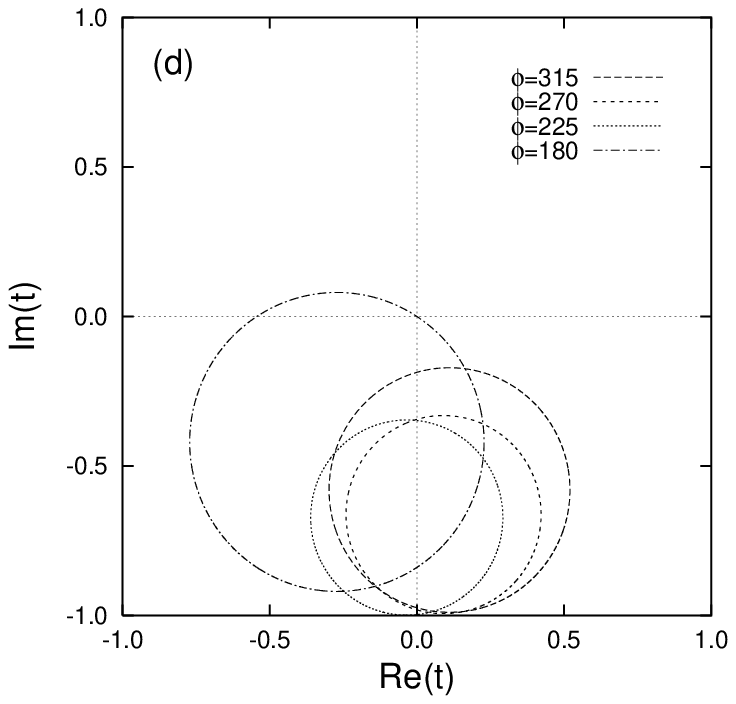}
 \includegraphics{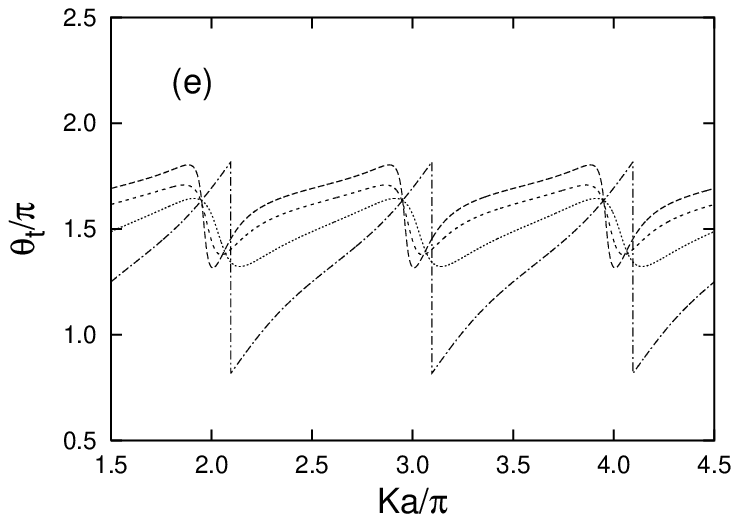}
 \includegraphics{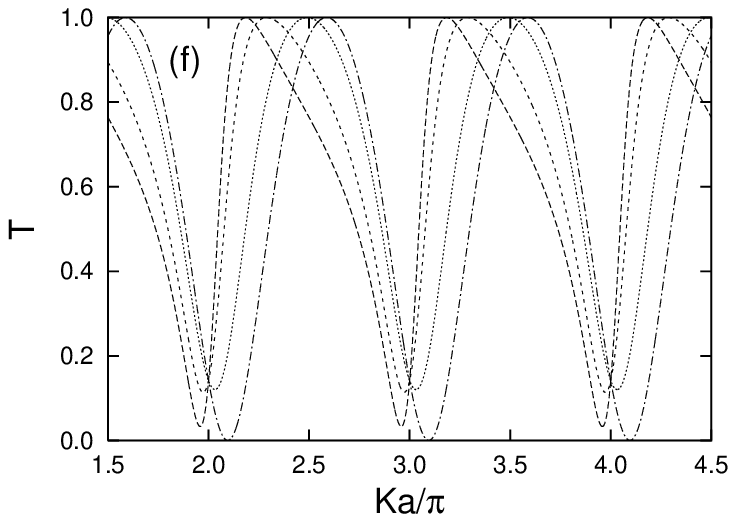}
\end{minipage} 
\vskip 0.5cm 
\caption{Behavior of the transmission amplitude $t$ with varying the AB phase $\phi$
for the AB ring with the $t$-stub. 
Panel descriptions are the same as in Fig.~\ref{ringdot}.
Model parameters are chosen as $k_FL = \pi/2$; 
$\epsilon_{L,R} = 1/2$, $\lambda_{L,R1}=-1$, and $\lambda_{L,R2} = 1$; 
$\epsilon_t = 4/9$, $\lambda_{t1} =-1$ and $\lambda_{t2} =1$; 
$T_L = T_R = 0.8$.  
\label{ringstub}}
\end{figure}
\begin{figure}
\includegraphics{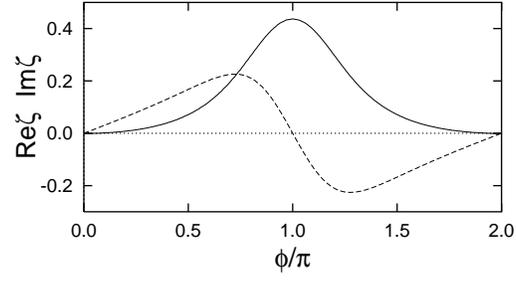}
\vskip 0.5cm
\caption{Transmission zeros as function of the AB phase $\phi$
in the AB ring with the $t$-stub system.  
Solid(dashed) line is the real(imaginary) 
part of the shifted zero $\zeta(\phi)$, respectively.    
\label{zerostub}}
\end{figure}

\end{widetext}

\end{document}